\documentclass[12pt]{article} 
\pdfoutput=1
\usepackage{epsfig}
\usepackage{graphicx}
\usepackage{jheppub}
\usepackage{amsmath}
\usepackage{amsfonts}
\usepackage{amssymb}
\usepackage{graphicx}
\usepackage{inputenc}
\usepackage{slashed}

\usepackage{color}
\usepackage{subcaption}
\def\beq{\begin{equation}}
\def\eeq{\end{equation}}
\def\beqa{\begin{eqnarray}}
\def\eeqa{\end{eqnarray}}

\renewcommand{\epsilon}{\varepsilon}

\def\ifm{\ifmmode}

\definecolor{darkgreen}{rgb}{0.0, 0.4, 0.13}

\newcommand{\ep}{\epsilon}

\newcommand{\zb}{\bar z}
\newcommand{\la}{\lambda}

\def\[{\begin{equation}}
\def\]{\end{equation}}

\def\f21{{}_2F_{1}}

\def\O{\mathcal{O}}

\newcommand\sss{\scriptscriptstyle}
\newcommand\as{\alpha_{\sss S}}

\def \al #1 {\frac {\as({#1})}{\pi} }

\allowdisplaybreaks[1]

\title{One-loop QCD contributions to differential cross-sections for Higgs production at N$^3$LO}

\author[a]{Charalampos Anastasiou,}
\author[a]{Caterina Specchia}
\affiliation[a]{Institute for Theoretical Physics, ETH Z\"urich, 8093 Z\"urich, Switzerland.}

\emailAdd{babis@phys.ethz.ch}
\emailAdd{specchic@phys.ethz.ch}

\abstract{
	We present one-loop contributions to the fully differential
        Higgs boson gluon-fusion cross-section
	for Higgs production via gluon fusion. Our results constitute a
        necessary ingredient  of a complete  N$^3$LO determination of
        the cross-section. We perform our computation using a
        subtraction method for the treatment of soft and collinear
        singularities.  We identify the infrared divergent parts 
	in terms of universal splitting and eikonal functions, and
        demonstrate how phase-space integrations yield poles (up to
        $1/\ep^6$ ) in the dimensional regulator $\ep=(4-d)/2$. 
       We compute the coefficients of the $\ep$ expansion, including
       the finite part numerically.  As a demonstration of our
       numerical implementation, we present the corrections at N$^3$LO
       due to one-loop amplitudes in the rapidity and transverse momentum of the Higgs boson.
}


\notoc
\begin{document}
\begin{flushright}
\vspace*{-25pt}
\end{flushright}
\maketitle
\allowdisplaybreaks

\section{Introduction} 

\label{RV2} 

Particle physics has entered an era of precision phenomenology
which, at its core, aims to probe the interactions of the Higgs boson
and other known or undiscovered particles.  
With the second run of the LHC,  the statistical accuracy of the experimental measurements
will increase significantly, allowing a precise determination of a
variety of differential cross-sections 
and kinematic distributions.
Precise theoretical predictions for fully differential
cross-sections are highly desired. Their comparison to the
measurements will offer valuable tests of the Standard Model and will
set constraints to physics beyond the Standard Model.

Recently, the inclusive Higgs boson cross-section 
was computed through
next-to-next-to-next-to-leading order (N$^3$LO) in
the strong coupling perturbative
expansion~\cite{Anastasiou:2015ema,Mistlberger:2018etf}. 
Important achievements have also been accomplished towards differential
Higgs cross-section at N$^3$LO.
The N$^3$LO  gluon-fusion Higgs production cross-section  with a jet-veto 
has been obtained~\cite{Banfi2016} by combining 
the fully differential cross-section for $pp \to H+1{\,\,\rm jet}$ at
NNLO~\cite{Chen2016,Boughezal2015a,Boughezal2015b,Caola2015,Chen2015} with the N$^3$LO inclusive cross-section~\cite{Anastasiou:2015ema,Mistlberger:2018etf}.
Other differential Higgs cross-sections have been computed by means of a threshold 
expansion~\cite{Dulat:2018bfe,Dulat:2017brz,Dulat:2017prg}
and  the $q_T$ subtraction formalism which is 
being extended to N$^3$LO~\cite{Cieri:2018oms}.

We envisage a calculation of the fully differential Higgs
cross-section with a direct subtraction of infrared
divergences from the phase-space integrations over the partonic
radiation associated with the Higgs boson production. At N$^3$LO, one
must add the fully differential partonic cross-sections for radiative processes of the type:
\begin{equation}
\mbox{parton} + \mbox{parton} \to \mbox{Higgs} + n \; \mbox{partons}, \quad n
\leq 3. 
\end{equation}
The complete set of phase-space integrations in the above has been
achieved inclusively, as it was required for the determination of the total 
Higgs cross-section.  For a fully differential Higgs cross-section, the integrals over the
phase-space must be performed as functionals of a generic
infrared-safe measurement function.  This can be achieved numerically,
after the subtraction and cancelation of soft/collinear divergences.  

Processes with one parton in the final state ($n=1$),
represent the simplest non-trivial case of phase-space integrations. 
At N$^3$LO,  they receive contributions from the ``real-virtual-virtual (RVV)'' 
interference of two-loop amplitudes and tree-level amplitudes
as well as the square of one-loop amplitudes (real-virtual-squared
(RV)$^2$). 
The RVV contributions to the fully differential Higgs boson
have been studied in Ref.~\cite{Duhr:2014nda}.

In this article, we consider the  (RV)$^2$ contributions, making a
modest step towards a complete determination of the fully
differential cross-section at N$^3$LO.  
In particular, we revisit the soft and collinear singular limits of
the  one-loop amplitudes in terms of universal splitting amplitudes
and soft-currents at one-loop. Then we isolate the singular terms of
the partonic cross-sections with the aid of appropriate counterterms. 
This allows us to compute the coefficients of the expansion in the
dimensional regulator $\ep= 2 -\frac d 2$
numerically for arbitrary measurement functions.  
As explicit examples, we present these coefficients differentially, in
bins of the Higgs transverse-momentum or its rapidity. 

\section{Setup}
\label{sec:gensetup}

We consider processes 
$$i(p_1) + j(p_2) \rightarrow h(p_h) + k(p_3)\,,$$
where $i,j,k$ denote quark, antiquark or gluon partonic flavors, 
$p_{1,2}$ are the momenta of the incoming partons,
$p_h$ is the momentum of the Higgs boson and
$p_3$ is the momentum of the radiated parton.
The Mandelstam variables are:
\begin{align}
	p_{1}^2=p_{2}^2=p_3^2&=0, &
	(p_1+p_2)^2&= s_{12},\\
	(p_2-p_3)^2&= s_{23}&
	(p_1-p_3)^2&= s_{13},
\end{align}
with  $ s_{12}+ s_{23}+ s_{13}=m_h^2$.
We parameterise the final-state momenta in terms of dimensionless
positive variables $z,\lambda \leq 1$ as in:
\begin{align} \label{eq:GHpar}
	p_3 &=\bar{z} \left( \lambda p_1 +\bar\lambda p_2
              +\sqrt{s_{12} \lambda \bar \lambda  }  \eta_{\perp} \right),
\end{align}
where
\[
	 p_{1,2} \cdot \eta_\perp=0, \quad \eta_\perp^2=-1 
\]
and
\[
 z=m_h^2/s_{12}, 
	\quad	\bar{z}= 1-z, \quad
	\bar\lambda=1-\lambda.
\]

We evaluate perturbatevely the amplitudes for the processes 
\begin{align*} 
	& g + g \rightarrow g+h \\
	& q + g \rightarrow q+h 
\end{align*}
(and the ones related to the above by crossing symmetry and/or 
charge conjugation)
in the Standard Model and in the limit of a very heavy top-quark.  
The leading contribution to this limit in the strong-coupling sector 
is described by the Lagrangian density: 
\[
	\mathcal{L} = \mathcal{L}_{QCD} - \frac{1}{4}C_1 G_{\mu\nu}G^{\mu\nu}h\,,
\]
where $\mathcal{L}_{QCD}$ is the QCD Lagrangian density (with  $n_f=5$
massless quark flavours and $N_c$ number of colours),
$h$ is the Higgs boson field,
$G_{\mu\nu}$ the gluonic field-strength tensor and
$C_1$ is the Wilson coefficient~\cite{Chetyrkin:1997un,Chetyrkin:2005ia,Schroder:2005hy}
that arises from matching the effective theory to the full Standard Model.

We renormalise the bare strong coupling constant $\alpha_s \equiv
\frac{g_s^2}{4 \pi}$ and the Wilson coefficient
in the $\overline{\rm MS}$ renormalisation scheme:
\[
	\alpha_s=\alpha_s(\mu) \left(\frac{\mu^2}{4\pi}\right)^{\epsilon}e^{\epsilon \gamma_E} Z_\alpha,
	\qquad
	C_1=C_1(\mu^2)Z_C,
\] 
where the multiplicative factors $Z_\alpha(\mu^2)$ and $Z_C(\mu^2)$ 
are given by:
\begin{align}
 Z_\alpha&=1
	-\frac{\beta_0}{\epsilon} \left(\frac{\alpha_s(\mu)}{\pi} \right)
	+\O(\alpha_s^3) \\
Z_C&=1
	-\frac{\beta_0}{\epsilon} \left(\frac{\alpha_s(\mu)}{\pi} \right)
	+\O(\alpha_s^2)
\end{align}
The renormalised Wilson coefficient is given by:
\begin{align} \label{eq:DefCW}
	C_1(\mu) &= - \frac{\alpha_s(\mu)}{3 \pi v} \Bigg\{ 1
                     +\left(\frac{\alpha_s(\mu)}{\pi}
                     \right)\,\frac{11}{4} 
	+ \mathcal{O}(\alpha_s^2) \Bigg\} .
\end{align}

We compute the required one-loop amplitudes as well as their soft/collinear limits 
in conventional dimensional regularisation (CDR).  
The ``form-factors''  of the amplitudes as computed in CDR
suffice to determine fully the amplitudes for the scattering of partons of
definite helicity.  The universal collinear 
and soft limits of helicity amplitudes are known in the
literature~\cite{Kosower:1999rx,Badger:2004uk,Catani:1998bh,Catani:2000pi,Duhr:2013msa}
 and we verify that our results agree with them.

\section{Tree and one-loop amplitudes}
In this section, we present the tree and one-loop amplitudes which are required
for the gluon-fusion Higgs production cross-section at N$^3$LO in perturbative
QCD.  

\subsection{The $gg \rightarrow h$ amplitude}
Let us start first with the gluon-gluon scattering  process
\[
	g(p_1)+g(p_2) \rightarrow h(p_h) \,,
\]
with $p_h=p_1+p_2$.
For physical external polarisations, 
$\epsilon(p) \cdot p=0$, 
we can write the amplitude as
\[
	M_{g_a g_b \rightarrow h} = i\,\delta^{ab}  \, \ep_{\mu_a}(p_1) \ep_{\mu_b}(p_2)\, A_h\, \left( p_1\cdot p_2 \, g^{\mu_a \mu_b} -p_1^{\mu_b}p_2^{\mu_a} \right) \,.
\]
The coefficient $A_h$ admits a perturbative expansion in the bare
strong coupling constant $\alpha_s$,
\[
	A_h=  C_1\, \left( A_h^{(0)}+\frac{\alpha_s}{\pi} 
	 \, A_h^{(1)} +O(\alpha_s^2) \right).
\]
In the following, we will be concerned only with the first two terms in the expansion,
which read
\begin{align}
A_h^{(0)}&=1\\
A_h^{(1)}&=(-s_{12})^{\ep}\frac{c_\Gamma\,(4\pi)^\ep}{2}
	\frac{(1-3 \epsilon +2 \epsilon^2+\epsilon^3)}{\epsilon^2 (1-\epsilon)(1-2 \epsilon)}\,.	
\end{align}
where 
\begin{equation}
c_\Gamma = \frac{\Gamma(1+\ep) \Gamma(1-\ep)^2}{ \Gamma(1-2\ep)}\, .
\end{equation}

\subsection{The $gg \rightarrow gh$ amplitude}

We now the consider the process
\[
	g(p_1)+g(p_2) \rightarrow h(p_h) + g(p_3),
\]
with $p_h=p_1+p_2+p_3$.
If we choose polarisation vectors $\epsilon^\mu_i \equiv
\epsilon^\mu(p_i)$  for the external gluons which satisfy
\[ \label{polnorm}
	\epsilon_1 \cdot p_2=0, \qquad \epsilon_2 \cdot p_3=0, \qquad \epsilon^*_3 \cdot p_1=0,
\]
we can cast  the amplitude 
in the form:
\begin{align} \label{mgghg}
	M_{gg \rightarrow gh} =&\frac{f^{a_1 a_2 a_3}}{s_{12} s_{23} s_{13}}
	\left[
	A_1
	\left( 
	 \epsilon_1 \cdot p_3 \,  \epsilon_2 \cdot p_1 \, \epsilon^*_3 \cdot p_1 
	\right) \right. \nonumber\\
	&+A_{2a}
	\left(
	s_{12}\, \epsilon_1 \cdot p_3 \,  \epsilon_2 \cdot \epsilon^*_3 -  \epsilon_1 \cdot p_3 \,  \epsilon_2 \cdot p_1 \, \epsilon^*_3 \cdot p_1
	\right)\nonumber\\
	&+A_{2b}
	\left(
	s_{23}\, \epsilon_2 \cdot p_1 \,  \epsilon_1 \cdot \epsilon^*_3 -  \epsilon_1 \cdot p_3 \,  \epsilon_2 \cdot p_1 \, \epsilon^*_3 \cdot p_1
	\right)\nonumber\\
	&+A_{2c}
	\left. \left(
	s_{13}\,  \epsilon^*_3 \cdot p_2 \,  \epsilon_1 \cdot \epsilon_2-  \epsilon_1 \cdot p_3 \,  \epsilon_2 \cdot p_1 \, \epsilon^*_3 \cdot p_1
	\right)
	\right].
\end{align}
Squaring and summing over polarisations and colour, we find 
\begin{equation}
\label{M2}
	\sum_{pols} \sum_{cols} |M_{gg\rightarrow gh}|^2 = 
	\frac{N_c \, (N_c^2-1)}{s_{12} s_{23} s_{13}}
	\left(
	|A_1|^2 + (d-3) \sum_{i=a,b,c} |A_{2i}|^2
	 \right).
\end{equation}
\\
We can further relate the form factors  $A_1, A_{2a}, A_{2b}, A_{2c}$ 
to helicity amplitudes~\cite{Gehrmann:2011aa}:
\begin{align}
	|M^{+++}_{gg \rightarrow gh}\rangle&= \alpha_{+++} \, \frac{1}{\sqrt{2}}\frac{m_h^4}{\langle p_1 p_2\rangle \, \langle p_2 p_3\rangle \, \langle p_1 p_3\rangle} \label{hel+++} \\
	|M^{++-}_{gg \rightarrow gh} \rangle&= \alpha_{++-} \, \frac{1}{\sqrt{2}} \frac{\left[ p_1 p_2\right]^3}{\left[ p_2 p_3\right] \, \left[ p_1 p_3\right] } \label{hel++-} \\
	|M^{+-+}_{gg \rightarrow gh} \rangle&= \alpha_{+-+} \, \frac{1}{\sqrt{2}} \frac{\left[ p_1 p_3\right]^3}{\left[ p_1 p_2\right] \, \left[ p_2 p_3\right] } \label{hel+-+} \\
	|M^{-++}_{gg \rightarrow gh} \rangle&= \alpha_{-++} \, \frac{1}{\sqrt{2}} \frac{\left[ p_2 p_3\right]^3}{\left[ p_3 p_1\right] \, \left[ p_1 p_2\right] } \label{hel-++},
\end{align}
where the coefficients $\alpha_i$'s
are related to the amplitude coefficients $A_i$'s via
\begin{align} \label{ab}
	\alpha_{+++}&=\frac{1}{2 m_h^4} \left( A_1+A_{2a}+A_{2b}+A_{2c} \right) \\
	\alpha_{++-} &= \frac{1}{2 s_{12}^2} \left( A_1+A_{2c}-A_{2a}-A_{2b} \right) \\
	\alpha_{+-+}&=\frac{1}{2 s_{13}^2} \left( A_1+A_{2b}-A_{2a}-A_{2c} \right) \\
	\alpha_{-++}&=\frac{1}{2 s_{23}^2} \left( A_1+A_{2a}-A_{2b}-A_{2c} \right).
\end{align}

Notice that all of the above relations are valid at any
order in the perturbative expansion in the strong coupling constant. 
We now expand the form-factors perturbatively:
\[
	A_i= \,C_1  \sqrt{4 \pi \alpha_s}  \left( A_i^{(0)} +
	 \frac{\alpha_s}{\pi}
	  \, A_i^{(1)} + \O(\alpha_s^2) \right).
\]

The leading order amplitude coefficients $A_i$ are
\begin{align}
	&A_{2a}^{(0)}=  \bar z (1-\lambda \bar z)  \\
	&A_{2b}^{(0)}= \bar z (1-\bar \lambda \bar z)   \\
	&A_{2c}^{(0)}=- (1-\lambda \bar z) (1-\bar \lambda \bar z)   \\
	&A_{1}^{(0)}=- \left[1-\bar z +(1-\lambda \bar \lambda)\bar
          z^2\right] \, , 
\end{align}
where the final-state momenta are given by Eq.~\eqref{eq:GHpar}.

At one-loop, the amplitude coefficients $A_{1}^{(1)} ,A_{2a}^{(1)},
A_{2b}^{(1)}, A_{2c}^{(1)}$ are linear combinations of 
the bubble and box integrals, which are defined as
\begin{align*}
	\text{Bub}(q^2) &= \int \frac{d^dk}{i\pi^{d/2}} \frac{1}{k^2 \,(k+q)^2}=
	\frac{c_\Gamma}{\epsilon^2 (1-2 \epsilon)} (-q^2)^{-\epsilon},\\
	 \text{Box}(s,t,u) &= 
	\int \frac{d^dk}{i\pi^{d/2}} \frac{1}{k^2 (k+q_1)^2 (k+q_1+q_2)^2 (k+q_1+q_2+q_3)^2 } \nonumber\\
	&=2  \frac{c_\Gamma}{\epsilon^2} \frac{1}{s\, t} 
	\left[  
	(-t)^{-\epsilon} \, _2F_1\left(1,-\epsilon,1-\epsilon,\frac{-u}{s} \right) \right. \nonumber\\
	 & \left. \qquad +(-s)^{-\epsilon} \, _2F_1\left( 1,-\epsilon,1-\epsilon,\frac{-u}{t} \right)-(-m^2)^{-\epsilon}\, _2F_1\left( 1,-\epsilon,1-\epsilon,\frac{-m^2 u}{s\,t} \right)
	\right].  
\end{align*}
In the above,  $(q_1+q_2)^2=s,  \, (q_2+q_3)^2=t, \,  (q_1+q_3)^2=u$,
$q_1^2=q_2^2=q_3^2=0$ and $s+t+u=m^2$.
The hypergeometric function can be expanded in $\epsilon$ in terms of polylogarithms, 
\begin{eqnarray}
_2F_1\left(1,-\epsilon,1-\epsilon, x\right) &=& 1 + \ep \log(1-x) - 
\sum_{n=2}^{\infty} \epsilon^n {\rm Li}_{n} (x).  
\end{eqnarray}

The arguments of the master integrals which appear in
the amplitudes are 
\begin{align} 
	 &\left\{ \text{Bub}(s_{12}), \, \text{Bub}(s_{23}),  \, \text{Bub}(s_{13}),  \,\text{Bub}(m_h^2),  \, \right.\\ \nonumber
	&\quad \left.
	\text{Box}(s_{12},s_{23},s_{13}), \, \text{Box}(s_{12},s_{13},s_{23}),  \, \text{Box}(s_{13},s_{23},s_{12}) \right\}. \label{MIs}
\end{align}

It will be convenient to choose a basis of master integrals where the
box integrals are defined in $d+2$ dimensions.  
To shift the dimension, we use the dimensional shift relation~\cite{Bern:1992em}
\begin{equation}
	\text{Box}^{d}(s,t,u) = \frac{2 (d-3)}{s t} 
	\left[  
	-u \, \text{Box}^{d+2}(s,t,u) 
	-\frac{2}{d-4} \left( \text{Bub}(s) +\text{Bub}(t) - \text{Bub}(m^2) \right)
	\right]\,.
\end{equation}
The expressions for the amplitudes $A_i$ written in terms of these Master Integrals
are given in the appendix \ref{appendix}.
In order to simplify the notation,  we set 
 $s_{12}=1$ in these expressions and the rest of this work. We also denote
\[
 (-s_{12})^{a \ep} \to (-\hat{\mathsf{1}})^{a \ep}
\]
for any integer $a$, bearing in mind that the quantity in the
parenthesis has a small negative imaginary part. 
The mass dimensions of any quantity can be recovered easily with dimensional analysis.

\subsection{The $qg \rightarrow qh$ and $q\bar q \rightarrow gh$ amplitudes}

The amplitudes for the  $q(p_1) +g(p_2) \rightarrow q(p_3) +h(p_h)$
and the  $q(p_1)+ \bar q(p_2) \rightarrow g(p_3) + h(p_h)$ processes are related by crossing symmetry. We will
therefore present here only the amplitude for the former. For physical
gluon polarisations, it takes the form
\begin{align}
\label{eq:Mqg_gen}
	M_{qg \rightarrow qh} &= 
	T^{a_2}_{j_1 j_3} \left[ 
	A_2 \left( 
	\bar u(p_3) \, \slashed p_2 \, u(p_1) \,p_1 \cdot \epsilon_2  -
	\bar u(p_3) \, \slashed \epsilon_2 \, u(p_1) \, p_1 \cdot p_2 
	\right) \right.  \nonumber \\
	&\left. -A_3 \left( 
	\bar u(p_3) \, \slashed p_2 \, u(p_1) \,p_3 \cdot \epsilon_2  -
	\bar u(p_3) \, \slashed \epsilon_2 \, u(p_1) \, p_2 \cdot p_3
	\right)
	\right].
\end{align}
Squaring and summing over spins, polarisations and colour, we find 
\begin{eqnarray}
\label{M2qg}
	\sum_{spin} \sum_{cols} |M_{qg\rightarrow qh}|^2 &=& 
\frac{1}{2} \left(N_c^2-1\right)s_{13}
	\left[
	(d-2)s_{12}^2 \, |A_2|^2
\right.  
\nonumber \\ && \hspace{-1.8cm}
\left. 	
-  (d-4) s_{12} s_{23}  \left( A_2 A_3^* + A_3 A_2^*\right)
	-  (d-2) s_{23}^2 |A_3|^2
	\right]
\end{eqnarray}
Each of the amplitude coefficients in Eq.~\eqref{eq:Mqg_gen} can be expanded as a power
series in the strong coupling constant
\[
	A_i=  C_1 \sqrt{4 \pi \alpha_s}  \left( A_i^{(0)} + \frac{\alpha_s}{\pi}
	\, A_i^{(1)} + \O(\alpha_s^2) \right).
\]
At leading order 
\[
	A_2^{(0)}=A_3^{(0)}=\frac{1}{s_{13}}.
\]
The one-loop $A_2^{(1)}, A_3^{(1)}$ amplitude coefficients are
presented in the appendix~\ref{appendix}.

\section{Infrared divergences of one-loop amplitudes}

The one-loop amplitudes of the previous section are divergent in $d=4$
dimensions, due to singularities when the momenta of two adjacent massless particles become collinear, 
or when a massless particle is soft.  These divergences cancel in a
complete hadronic cross-section computation.  In the singular limits,
the amplitudes exhibit universal factorisation properties. We will
exploit them in order to isolate the divergent parts and  to
facilitate the integration of the one-loop contributions to the
partonic cross-sections, which we are computing here, into a future
complete hadronic cross-section computation.  

In the following subsections, we will recall the factorization of the
amplitudes in the limits where two external partons become collinear
or an external parton becomes soft.

\subsection{Collinear limits}
\label{sec:collLim}

In the limit where two external-particles become colliner, colour
ordered amplitudes  factorise in a universal
way~\cite{Bern:1998vu,Bern:1999ry,Bern:1995ix,Bern:1994zx,Kosower:1999rx,Badger:2004uk,Bern:2004cz,Gehrmann:2011aa,Duhr:2014nda}. In
this section we will compute explicitly the collinear limits of the $gg \to gh$
and $gq \to g h$ amplitudes
and cast them in terms of universal functions related to the tree and one-loop splitting amplitudes.   

\subsubsection{Collinear limits for the $gg \rightarrow gh$ amplitude}

Let us first consider the limit for $p_3 || p_2$ becoming collinear  
(the other limit $p_3 || p_1$ can be derived from this by symmetry $p_2 \leftrightarrow p_1$,
that is $\lambda \leftrightarrow 1-\lambda$).

At leading order, we find that 
\begin{align}
\lim_{\lambda \to 0} A_{2a}^{(0)}=& \bar z \\
\lim_{\lambda \to 0} A_{2b}^{(0)}=& \bar z (1-\bar z) \\
\lim_{\lambda \to 0} A_{2c}^{(0)}=&- (1-\bar z) \\
\lim_{\lambda \to 0} A_{1}^{(0)}=&- (1-\bar z +\bar z^2).
\end{align}
At one-loop, we find the following collinear limits for 
the amplitude coefficients, 
\begin{align}
 \lim_{\lambda \to 0} A_{2a}^{(1)} =&\bar z\, 
 	N_c \, 
	\left[ 
	(\lambda\, \bar z)^{-\epsilon}
	\left(
	\tilde{r}_1(w(\bar z))+\tilde{r}_2 
	\right)A_h^{(0)}
	-(1-\bar z)^{-\epsilon} \,A_h^{(1)}\right] \\
 \lim_{\lambda \to 0} A_{2b}^{(1)} =&\bar z(1-\bar z)\, 
 	N_c\,\left[
	 (\lambda\, \bar z)^{-\epsilon}
	\tilde{r}_1(w(\bar z))A_h^{(0)}
	-(1-\bar z)^{-\epsilon} \,A_h^{(1)}
	\right] \\
 \lim_{\lambda \to 0} A_{2c}^{(1)} =&-(1-\bar z)\, 
 	N_c\,\left[
	 (\lambda\, \bar z)^{-\epsilon}
	\tilde{r}_1(w(\bar z))A_h^{(0)}
	-(1-\bar z)^{-\epsilon} \,A_h^{(1)}
	\right] \\
\lim_{\lambda \to 0} A_1^{(1)} \,=&\,
 	N_c \, \left[
	(\lambda\, \bar z)^{-\epsilon}
	\left(
	-(1-\bar z +\bar z^2) \tilde{r}_1(w(\bar z))
	+ \bar z \, \tilde{r}_2
	\right)A_h^{(0)} \right.
\nonumber \\ 
&  \left. 
	+(1-\bar z +\bar z^2) (1-\bar z)^{-\epsilon} \,A_h^{(1)}
	 \right]
\end{align}
where the universal functions $\tilde{r}_1$ and $\tilde{r}_2$
are given by
\begin{align}
	\tilde r_1(w) &=2 \pi^2 
		\left[ 
		w f_1(w) + (1-w) f_1(1-w) -2 f_2
		\right], \\
	\tilde r_2 &=-4 \pi^2\frac{\ep^2}{(1-2\ep)(3-2\ep)}
		\left[ 1
		-\frac{1}{(1-\ep)} \, \frac{n_f}{N_c}
		\right]
		f_2 
\end{align}
with
\[
	w(\bar z)=\frac{\bar z}{\bar z-1}
\]
and the functions $f_1$ and $f_2$ are defined as
\begin{align}
	 f_1(w)&= \frac{2}{\ep^2} \frac{c_\Gamma}{(4 \pi)^{2-\ep}}
		\left[
		-\Gamma(1-\ep)\,\Gamma(1+\ep)
                 \frac{(1-w)^\ep}{w^{1+\ep}} -\frac{1}{w} 
\right. \nonumber \\ & \left. \hspace{1cm}
		+\frac{(1-w)^\ep}{w}\, _2F_1(\ep, \ep, 1+\ep; w)
		\right] \\
	f_2 &=- \frac{1}{\ep^2}\frac{c_\Gamma}{(4 \pi)^{2-\ep}}
\end{align}
Our results are in agreement with Ref.~\cite{Kosower:1999rx, Badger:2004uk}.

\subsubsection{Collinear limits for the $qg \rightarrow q h$  and 
$q \bar q \rightarrow g h$ amplitudes}

The collinear limit $p_3 || p_1$ is common to both 
one-loop amplitude coefficients $A_2^{(1)}$ and $A_3^{(1)}$ 
for the $qg \rightarrow q h$ amplitude.  Specifically, we find 
\begin{equation}
\lim_{\lambda \to 1} A_{2}^{(1)} = 
\lim_{\lambda \to 1} A_{3}^{(1)}=
N_c \, A_{2,3}^{(1),\text{LC}} +  \frac{1}{N_c} \, 
A_{2,3}^{(1),\text{SC}}
+n_f \, A_{2,3}^{(1),n_f}, 
\end{equation}
where
\begin{align}
	A_{2,3}^{(1),\text{LC}}&=
  \frac{1}{\bar\lambda \bar z}
	\left[ 
	\left(\bar\lambda \bar z \right)^{-\ep}
	\left(
	\tilde r_1(w(\bar z))+ c_\Gamma \frac{3(2-\ep)}{2\ep^2(1-2 \ep)(3-2 \ep)}
	\right) A_h^{(0)}
	-(1-\bar z)^{-\ep} A_h^{(1)}
	\right]\\
	A_{2,3}^{(1),\text{SC}}&=
	 \left(\bar\lambda \bar z \right)^{-1-\ep}
	c_\Gamma
	\frac{2(2 - \ep + 2 \ep^2)}{\ep^2 (1-2\ep)} 
	A_h^{(0)}\\
	A_{2,3}^{(1),n_f}&=-
	\left(\bar\lambda \bar z \right)^{-1-\ep}
	c_\Gamma
	\frac{2(1-\ep)}{\ep(1-2 \ep)(3-2\ep)}
	A_h^{(0)} 
\, . 
\end{align}
The above results are in agreement with Ref~\cite{Kosower:1999rx,Badger:2004uk}.
The other limit for $p_2 || p_3$  ($\lambda \to 0$) is not singular.  
Similarly, the one-loop amplitudes for the process $q \bar q \to g h$ 
are not singular in the above collinear limits.

\subsection{Soft limit}
\label{sec:softLim}

We now turn our attention to  the factorisation of the amplitudes in
their soft limits.  

At tree-level, the soft limit, $\bar z \to 0$,  of the $gg \to gh$ amplitude is 
\begin{align}
\lim_{\bar z \to 0} A_{2c}^{(0)}=&\lim_{\bar z \to 0} A_{1}^{(0)}=- 1 \\
\lim_{\bar z \to 0} A_{2a}^{(0)}=& \lim_{\bar z \to 0} A_{2b}^{(0)}= 0,
\end{align}
and, at one-loop, 
\begin{align}
\lim_{\bar z \to 0} A_{2c}^{(1)} =& \lim_{\bar z \to 0} A_1^{(1)} =
	 \,\left[
	N_c \,A_h^{(1)} -
	\frac{(4\pi)^\ep}{4}
	 (-\hat{\mathsf{1}})^{-\epsilon}(\lambda \, \bar \lambda \, \bar z^2)^{-\epsilon} c_\Gamma\, \eta^{(1)}_{\text{soft}}A_h^{(0)}
	\right]\\
\lim_{\bar z \to 0} A_{2a}^{(1)} =&\lim_{\bar z \to 0} A_{2b}^{(1)} =0,
\end{align}
with 
\begin{equation}
	\eta^{(1)}_{\text{soft}}=-N_c\,\frac{\Gamma(1-\epsilon)\Gamma(1+\epsilon)}{\epsilon^2}.
\end{equation}
The above results are in agreement with Ref.~\cite{Badger:2004uk,Duhr:2013msa}. 

The one-loop amplitudes for both $qg \to qh$ and $q\bar
q \to gh $ processes are not singular in the soft limit.

\section{Hadronic cross-section and subtraction of infrared divergences}
\label{sec:Hadrxs}

We now consider the hadronic production of a Higgs boson in
association with a parton $i$ in the final state, 
\[ 
{\rm Proton}\left(P_1\right) + {\rm Proton}\left(P_2\right) 
\to h(p_h) + i(p_3).   
\]
The proton momenta $P_i$ in the hadronic centre of mass frame are given by
\begin{equation}
	P_1 = \frac{\sqrt S}{2} \left(1, 0, 0,  1 \right), \quad P_2 = \frac{\sqrt S}{2} \left(1, 0, 0, -1 \right),
\end{equation}
with $\sqrt{S}$ being the collider centre of mass energy.   
The  hadronic cross-section is given by the integral 
\begin{align}
	\sigma(\tau, \O)=&\tau \sum_{ij} \int_{\tau}^1 \frac{dz}{z}\, \int_{\tau/z}^1 \frac{dx_1}{x_1} \int_0^1 d\lambda\, \int_0^{2\pi} \frac{d\phi}{2\pi}
	f_i(x_1) \, f_j\left( \frac{\tau}{x_1 z}\right) 
	\frac{1}{z}\frac{d\hat\sigma_{ij}}{d\lambda}(z, \lambda, m_h^2) 
	\mathcal{J}_\O(z, \lambda, m_h^2)\,,
\end{align}
where 
\[
	\tau \equiv \frac{m_h^2}{S}\,,
\]
$x_1$ and $x_2$ are the Bjorken fractions
\[
	p_i=x_i \, P_i, \quad i=1,2 \; \, 
\]
and $f_i(x_i,\mu_F)$ are the renormalised parton distribution functions. 
The sum runs over all the initial-state parton pairs. 
For the computation of the fully differential cross-section 
at N$^3$LO, we need, among other contributions,  to integrate the matrix-elements of
Eq.~\eqref{M2}  and Eq.~\eqref{M2qg} over the phase-space of the
final-state partons, weighted with an infrared-safe measurement
function $\mathcal{J}_\O(p_h)$.  The matrix-elements are independent
of the azimuthal angle $\phi$. 

The momentum of the Higgs boson can be
conveniently parametrised in terms of its
 rapidity $Y_h$ and transverse momentum $p_T$,
defined in the plane transverse to the beam axis as
\[
	Y_h=\frac{1}{2} \log \left( \frac{E+p_z}{E-p_z} \right),
	\qquad
	p_T=\sqrt{E^2-p_z^2-m_h^2}.
\]
In terms the partonic variables of Eq.~\eqref{eq:GHpar},
these variables can be rewritten as
\begin{equation}
	Y_h=\frac{1}{2} \log \left[
	\frac{x_1}{x_2}
	\frac{1-\bar z \lambda}{1-\bar z \bar \lambda}
	\right],
	\qquad
	p_T^2 = x_1 x_2 S \,\bar z^2  \lambda  \bar \lambda.
\end{equation}

The partonic cross-section $\hat\sigma_{ij}$, differential in the variable $\lambda$,
for the different initial state
parton contributions is given by
\[
	\frac{d\hat\sigma_{ij}}{d\lambda}(z) = \frac{1}{2s_{12}}\int d\Phi_{h+1} \delta(\lambda'-\lambda)\sum_{pols} \sum_{cols}|M_{ij}|^2(z, \lambda',m_h^2)\,,
\]
where $M_{ij}$ are the corresponding matrix elements, 
while
the phase space for the production of the Higgs 
plus one final state parton is given by 
\[
	d \Phi_{h+1} = \frac{s_{12}^{\frac{d}{2}-2}}{4 (2\pi)^{d-2}} \bar z^{d-3} d\Omega_{d-2} d\lambda'\,
	(\lambda'\bar\lambda')^{\frac{d}{2}-2}
	\theta(s_{12}) \theta(\bar{z}) \theta(\lambda') \theta(\bar\lambda').
\]

In order to be able to compute a finite differential cross-section,
we need to subtract the collinear and soft singularities that
arise in the corresponding limiting kinematics.
More specifically, 
we find that all the partonic cross-sections $\hat\sigma_{ij}$ 
have only single poles in the variables $z$ and $\lambda$
of the form 
\[
	\bar z^{-1+a_z \ep}, \,\lambda^{-1+a_\lambda \ep},\, \bar \lambda^{-1+a_{\bar \lambda} \ep}
\]
where the coefficients $a_i$ are integer numbers.
Therefore, we can regulate the divergencies 
by subtracting the corresponding limiting contribution
at the integrand level and adding back the integrated
counterterm, as shown schematically in the following example
for a general singular variable $x$,
\begin{align}
\label{eq:plus}
	\int_0^1 dx \frac{f(x)}{x^{1+a \ep}}&=\int_0^1 dx \frac{f(x)-f(0)}{x^{1+a \ep}} +f(0)\,\int_0^1 dx \frac{1}{x^{1+a \ep}} \nonumber\\	
			&=  \sum_{n=0}^\infty 
			\frac{(-a \ep)^n}{n!} \int_0^1 dx \, \log^n(x) \frac{f(x)-f(0)}{x} + \frac{f(0)}{(-a\ep)}\,.
\end{align}

For the $gg$ initial states,
the collinear limiting behaviour for the matrix elements squared
can be cast in the following form
in terms of the universal functions $r_1$ and $r_2$,
\begin{align}
 \sum_{pols} \sum_{cols}|M^{(1)}_{gg}|^2&(z,\lambda\to 0,m_h^2) =N_{gg} \,C_1^2\, \frac{\alpha_s^3 \, e^{3 \ep\,\gamma_E }}{(4 \pi)^{3 - 3 \ep}\,16 \pi^4 \la \bar z^2} \quad \times \\
 	&\,\left\{
 	c_1(\bar z,\epsilon)  
 	\left[
 	\left|  r_1\, A_h^{(0)} \right|^2
	+\left| \tilde{A}_h^{(1)} \right|^2 
	-2 \mathcal{R}e \left( r_1 \, A_h^{(0)} \,\tilde{A}_h^{(1)*} \right)
	\right]  \right.\nonumber\\
 	&\left.
	+ c_2(\bar z,\epsilon)\,\left|  r_2\, A_h^{(0)} \right|^2 
	+c_3(\bar z,\epsilon) \mathcal{R}e\left[
	  r_1\, r_2^* \left|A_h^{(0)} \right|^2 
	-\left( r_2 A_h^{(0)} \,\tilde{A}_h^{(1)*}\right)
	\right]
	\right\}\nonumber
\end{align}
with
\begin{align}
&\tilde{A}_h^{(1)}=(1-\bar z)^{-\epsilon} A_h^{(1)}, \qquad r_i =(\lambda\, \bar z)^{-\epsilon} \tilde r_i \, \quad i=1,2\\ 
& c_1(\bar z,\epsilon)= 2(1-\epsilon)\left(1-\bar z+\bar z^2 \right)^2 \\
& c_2(\bar z,\epsilon)= 2(1-\epsilon) \bar z^2 \\
& c_3(\bar z,\epsilon)= -2\,\bar z \left( 1-2 (1-\epsilon) \bar z +\bar z^2 \right)
\end{align}
and similarly for $ \sum_{pols} \sum_{cols}|M^{(1)}_{gg}|^2(z, \lambda\to 1,m_h^2)$, with the exchange $\lambda \leftrightarrow \bar \lambda$.
Here $N_{gg}$ is the initial averaging factor over the spins and colours,
\[
	N_{gg}=\frac{1}{4(1-\epsilon)^2(N_c^2-1)^2}\,,
\]
and we have already normalised the coupling constant
with the factor
$
 \frac{e^{\epsilon \gamma_E}}{(4\pi)^{\epsilon} }\,.
$
The soft subtraction term, instead, is given by
\begin{align}
	 \sum_{pols} \sum_{cols}|M^{(1)}_{gg}|^2&(z\to1,\lambda,m_h^2)=N_{gg} \,C_1^2\,\frac{\alpha_s^3 \,e^{3 \ep \,\gamma_E}}{(4 \pi)^{3 - 3 \ep}}  \,
 	\frac{2(1-\epsilon)}{16 \pi^4 \la\bar \la \bar z^2} \quad\times\nonumber\\
	&\left[
	\left|A_h^{(1)} \right|^2 
	+ \left| \tilde{r}^{(1)}_{\text{soft}}\, A_h^{(0)} \right|^2
	- 2 \mathcal{R}e\left(A_h^{(1)} \,\tilde{r}^{(1)}_{\text{soft}}\, A_h^{(0)*}\right)
	\right]
\end{align}
where
\begin{align}
	&  \tilde{r}^{(1)}_{\text{soft}} =\frac{(4 \pi)^{\ep}}{4}
	(-\hat{\mathsf{1}})^{-\epsilon}(\lambda \, \bar \lambda \, \bar z^2)^{-\epsilon} c_\Gamma\, \eta^{(1)}_{\text{soft}}.
\end{align}

For the $qg$ initial states, the matrix elements squared
in the collinear limit $\lambda \rightarrow 1$
take the form
\begin{align}
	\sum_{pols} \sum_{cols}|M^{(1)}_{qg \to qh}|^2&(z, \lambda\to 1,m_h^2) =
		N_{qg} \,C_1^2\, \frac{\alpha_s^3 \,e^{3 \ep\,\gamma_E}}{(4 \pi)^{3 - 3 \ep}} \frac{(N_c^2-1)}{ 2} \,
		\frac{(1-\ep (1-\bar z)^2 +\bar z^2)}{16 \pi^4}\bar \lambda \bar z
		\quad \times
		\nonumber\\
		&
		\qquad\qquad\qquad\qquad\qquad
		\left|
		N_c \, A_{2,3}^{(1),LC} + \frac{1}{N_c} \,
                  A_{2,3}^{(1),SC} + n_f \, A_{2,3}^{(1),n_f}
		\right|^2
\end{align}
with the initial averaging factor over the spins and colours,
\[
	N_{qg}=\frac{1}{4(1-\epsilon)(N_c^2-1)N_c}\,.
\]
We recall that the partonic cross-section for the $q\bar q$ initial states
does not present any collinear nor soft singularity, and hence no 
subtraction is needed for this contribution.

With the aid of Eq.~\eqref{eq:plus} and the above infrared limits, we
can expand the partonic cross-sections in the dimensional regulator
$\epsilon$.  The resulting expressions are lengthy but straightforward
to derive and we refrain from presenting them here.   Our results are in agreement with an independent calculation
performed for the purposes of Ref.~\cite{Dulat:2018bfe}.

\section{Numerical Results}
\label{sec:Numerics}

We implement the results for the integrated counterterms and 
the finite part of the unrenormalized cross-section 
in a private numerical code
and compute distributions for the Higgs rapidity and transverse momentum.
Note that in the subtraction procedure we adopt to remove the soft and collinear
divergences, we need to add back integrated counterterms 
with explicit poles up to $\ep^{-6}$.

We consider a centre of mass energy of 13 TeV.
We use the NNLO PDF4LHC15 set~\cite{Harland-Lang2014} 
for the parton distribution functions,
as available from \texttt{LHAPDF}~\cite{Buckley:2014ana}
and evolve the strong coupling constant $\alpha_s$ to NNLO.
In addition, we set the Higgs boson mass to the value of 
$m_h=125$ GeV and take the same value of $\mu=m_h$ for
the renormalisation and factorisation 
scales.

The distributions we obtain, albeit non-physical as
they correspond only to a part of the complete N$^3$LO
calculation, serve as a validation of our results and of the
subtraction method employed to remove the infrared divergences.
In particular, this procedure provides a stable and reliable numerical
implementation.
\\

\begin{figure}[h]
	\centering
	\includegraphics[scale=0.8]{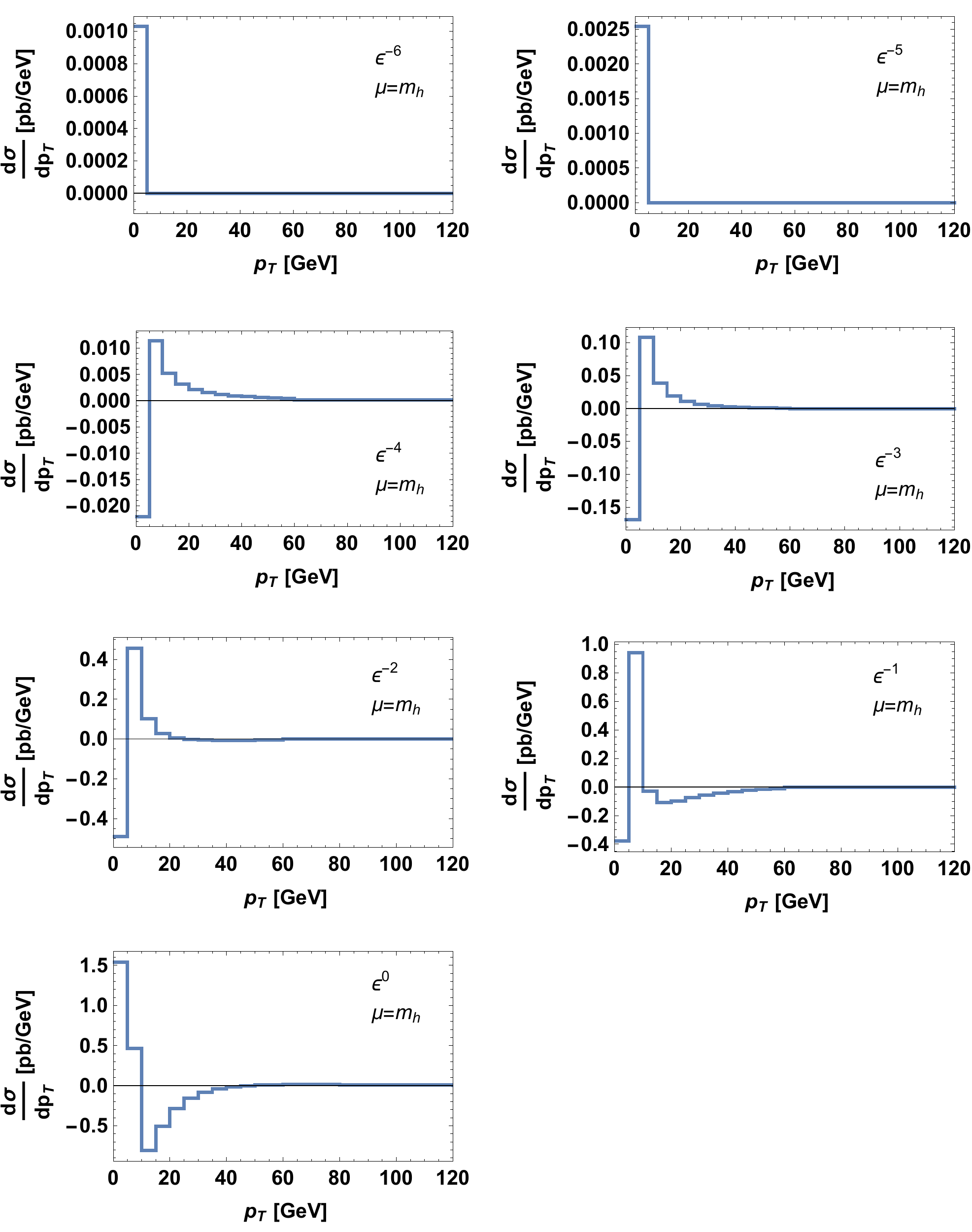}
	\caption{
	Distributions for the Higgs transverse momentum $p_T$ for each pole in the dimensional regulator $\ep$.
	The deepest poles $\ep^{-6}$ and $\ep^{-5}$ are proportional to the born cross-section,
	for which the expected distribution is obtained.
	The bands in the bins correspond to numerical uncertainties.
	}\label{fig:ptallep}
\end{figure}

In figure \eqref{fig:ptallep} we show
the distributions for the transverse momentum of the Higgs boson
$p_T$ for all the six $\ep$ poles and the finite part of the cross-section.
In the ones corresponding to the deepest poles $\ep^{-6}$ and $\ep^{-5}$
we recognise the expected behaviour, as these contributions are proportional
to the born cross-section.
For what concerns the other poles, 
as there is not a direct interpretation in terms of 
specific kinematic configurations,
the behaviour is anyway unphysical, with rather big cancellations
occurring between the first bins.
Nevertheless, we have checked that the sum of all the bins returns the total cross-section
for the corresponding pole in the dimensional regulator~$\ep$.

\begin{figure}[h]
	\centering
	\includegraphics[scale=0.8]{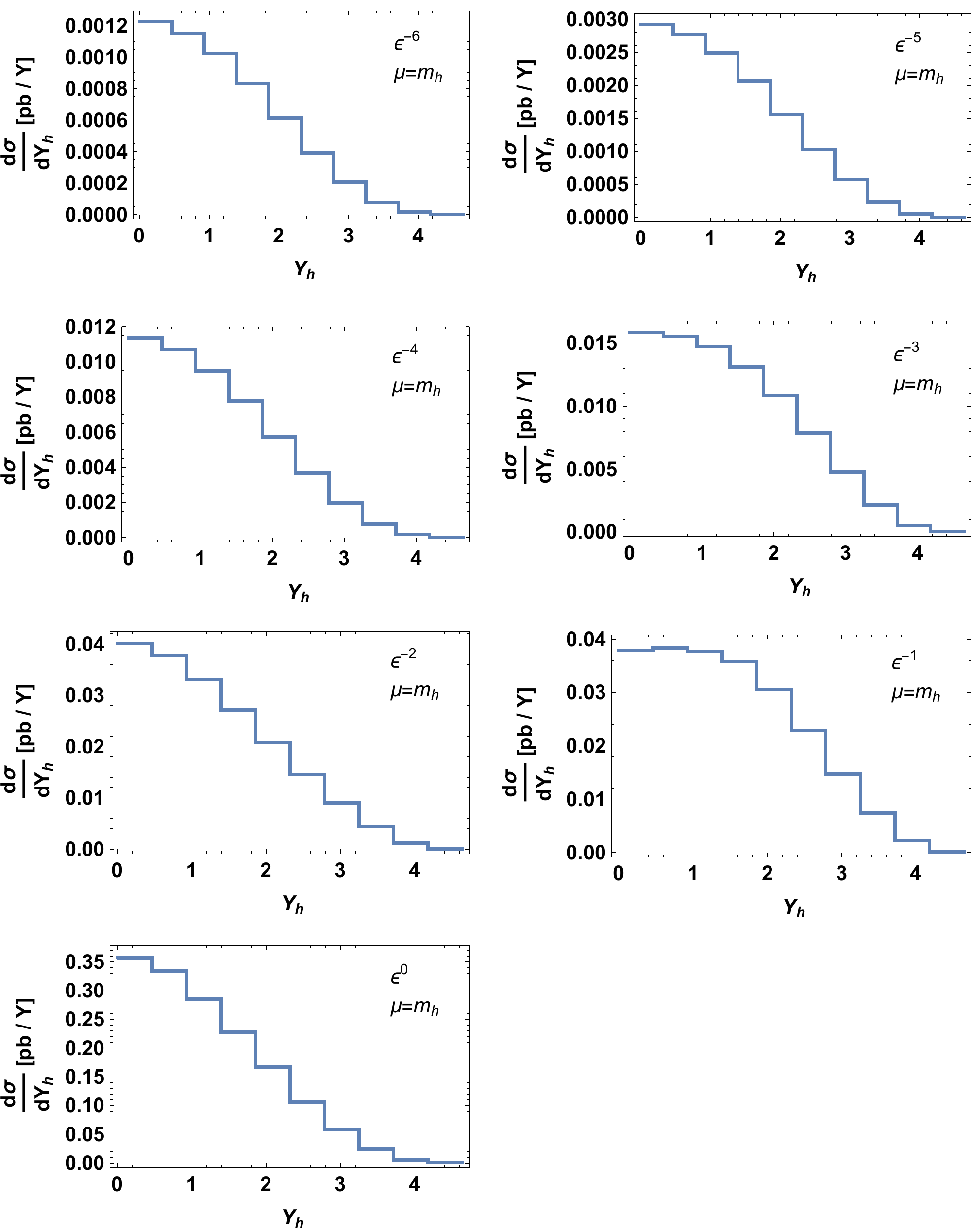}
	\caption{
	Distributions for the Higgs rapidity $Y_h$ for each pole in the dimensional regulator $\ep$.
	The bands in the bins correspond to numerical uncertainties.
	}\label{fig:yhallep}
\end{figure}

In figure \eqref{fig:yhallep} we show the rapidity distribution
for all the six $\ep$ poles and the finite part of the cross-section
for positive values of the rapidity $Y_h$,
since it is symmetric in $Y_h \leftrightarrow -Y_h$.

\clearpage

\newpage
\section{Conclusions}

We have presented the computation 
of the real-virtual squared contributions to the
fully differential cross-section for Higgs production via gluon fusion.
This is a part of the complete N$^3$LO fully differential
cross-section. 

One of the main results of our computation
is the analytic expression for all the soft and
collinear counterterms needed for the subtraction of the infrared
divergencies. In particular, we have been able to 
express these counterterms in terms of universal functions 
related to the QCD soft currents and splitting amplitudes at one loop.
Therefore, the results obtained here can be directly used in other
processes containing  a colourless final state particle and allow
for an easier and general numerical implementation
of differential cross-sections for such processes.

In order to validate our results and the subtraction procedure employed,
we implemented the contributions corresponding to 
the six poles in the dimensional regulator $\ep$ of the integrated
counterterms and the finite part of the cross-section in 
a  numerical code and obtained distributions for
the transverse momentum and rapidity of the Higgs boson.
These results, although non-physical, 
reproduced the expected behaviour for the 
deepest poles $\ep^{-6}$ and $\ep^{-5}$ corresponding
to born kinematic configurations,
while providing a stable numerical evaluation
for all the higher terms through the finite part of the $\epsilon$
expansion.

\section*{Acknowledgements}
We would like to thank V.~Del Duca,  A.~Lazopoulos, S.~Lionetti, F.~Moriello and A.~Pelloni for fruitful
discussions and B.~Mistlberger for comparisons before the publication of Ref.~\cite{Dulat:2018bfe}.
This project has received funding from the ETH Grant ETH-21 14-1 and
the pertQCD Advanced ERC Grant.

\appendix

\section{One-loop amplitudes}
\label{appendix}

Here we present the expressions for all the one-loop amplitude coefficients
for the processes $gg \to gh$, $qg \to qh$ and $q\bar q \to gh$
in terms of master integrals. 
The coefficient $c_\ep$ is defined as
\[
	c_\ep = \frac{(4 \pi)^\ep}{4} \,.
\]

\subsection{Amplitude coefficients for $gg \to gh$}
\begin{align*}
A_{2a}^{(1)} &= c_\ep \bigg\{ N_c \zb \left\{ -\left[\la \zb (1 - \la \zb) +    \ep \left(1 - \zb + 2 \la^2 \zb^2 - \la (1 + \zb)\right)\right] \text{Box}^{d+2}(s_{12},s_{13},s_{23}) \right.\\[5pt]
			& +\left[-(1 - \la) \zb (1 - \la \zb) + \ep \left(\la + 2 \zb - 3 \la \zb - 2 \la \zb^2 + 2 \la^2 \zb^2\right)\right] \text{Box}^{d+2}(s_{12},s_{23},s_{13}) \\[5pt]
			&\left.- \left[-1 + \la \zb +  \ep \left(2 + \la^2 (-1 + \zb) \zb - \la \zb (1 + \zb)\right)\right] \text{Box}^{d+2}(s_{13},s_{23},s_{12}) \right\}\\
			&+ N_c  \frac{-\zb (1 - \la \zb) + 3 \ep \zb (1 - \la \zb) +   \ep^2 \left[1 - 3 \zb - 2 \la (1 - \zb - \zb^2)\right]}{(1 - \ep) \ep} \text{Bub}(s_{12}) \\
			&+ N_c  \frac{\ep }{(1 - \ep) (1 - (1 - \la) \zb)^2 (1 - \la \zb)}\text{Bub}(m_h^2)\times\\
			&\quad\left[(1 - \zb) (-2 \la^4 \zb^3 (1 - \ep \zb) - (1 - \zb) (1 - 2 (1 - \ep) \zb + \zb^2)-  \right.\\
			&\quad \la^3 \zb^2 (2 - 5 \zb - (1 - 4 \ep) \zb^2)
			+ \la^2 \zb (2 - (5 + 2 \ep) \zb + (1 + 4 \ep) \zb^2 - 2 (1 - \ep) \zb^3) + \\
			&\,\,\, \left.\la (2 - 7 \zb + (11 + 2 \ep) \zb^2 - (5 + 4 \ep) \zb^3 + \zb^4))\right]  \\
			&- N_c  \frac{\zb \left[(1 - \la \zb)^2 - 3 \ep (1 - \la \zb)^2 + \ep^2 (3 - \zb - \la^2 (1 - 3 \zb) \zb - \la (1 + \zb)^2)\right]}{(1 - \ep) \ep (1 - \la \zb)} \text{Bub}(s_{13})\\
			&-\frac{\zb  }{(3 - 2 \ep) (1 - \ep) \ep (1 - (1 - \la) \zb)^2} \text{Bub}(s_{23}) \times \\
			&\quad\left[
			((1 - \la) n_f (\ep - \ep (1 - \la) \zb)^2 +  
			N_c \left(3 (1 - (1 - \la) \zb)^2 (1 - \la \zb) -  \right.\right.\\
			&\quad  11 \ep (1 - (1 - \la) \zb)^2 (1 - \la \zb)
			 + \ep^3 (-3 (1 - \zb)^2 + \la^3 \zb^2 (1 + 2 \zb) +  \la^2 \zb (6 - 5 \zb - 4 \zb^2) + \\
			 &\quad\la (1 - 10 \zb + 7 \zb^2 + 2 \zb^3))- 
			 \ep^2 (-11 (1 - \zb)^2 + \la^3 \zb^2 (2 + 9 \zb) +  \la^2 \zb (10 - 3 \zb - 18 \zb^2) +\\
			& \,\,\left. \left. \la (2 - 23 \zb + 12 \zb^2 + 9 \zb^3)))\right) \right] \bigg\}
\end{align*}

\begin{align*}
A_{2b}^{(1)} &=c_\ep \bigg\{  N_c \zb\left\{ 
			\text{Box}^{d+2}(s_{12},s_{13},s_{23}) \times\right. \\
			&\quad\left[-\la \zb (1 - \zb + \la \zb) +   \ep (1 - \zb + 2 \la^2 \zb^2 - \la (1 - 3 \zb + 2 \zb^2))\right]\\[5pt]
			&- \text{Box}^{d+2}(s_{12},s_{23},s_{13}) \times \\
			&\quad\left[(1 - \la) \zb (1 - (1 - \la) \zb) + \ep (-2 (1 - \zb) \zb + 2 \la^2 \zb^2 + \la (1 + \zb - 4 \zb^2))\right]\\[5pt]
			&+\text{Box}^{d+2}(s_{13},s_{23},s_{12})\left.\left[1 - \zb + \la \zb - \ep (2 - (2 - 3 \la + \la^2) \zb + (-1 + \la) \la \zb^2)\right]  \right\}\\
			&-\frac{N_c}{(1 - \ep) \ep} \text{Bub}(s_{12}) \times \\
			&\quad \left[\zb (1 - (1 - \la) \zb) - 3 \ep \zb (1 - (1 - \la) \zb) +    \ep^2 (1 + \zb - 2 \zb^2 - 2 \la (1 - \zb - \zb^2))\right]  \\[5pt]
			&+ N_c \frac{\ep(1 - \zb)}{(1 - \ep) (1 - (1 - \la) \zb) (1 - \la \zb)^2}  \,  \text{Bub}(m_h^2)\times\\[5pt]
			 &\quad\left[(1 - \zb) (1 - \zb - 2 \ep \zb) - 2 \la^4 \zb^3 (1 - \ep \zb) + 
			  \la^3 \zb^2 (2 + 3 \zb - (1 + 4 \ep) \zb^2) - \right.\\
			 &\quad \left. \la (2 - 3 \zb - (5 + 2 \ep) \zb^2 + 4 (1 + \ep) \zb^3) + 
			  \la^2 \zb (2 - (11 + 2 \ep) \zb + 4 (1 + \ep) \zb^2 + (1 + 2 \ep) \zb^3)\right]\\[5pt]
			& + \frac{\zb}{(3 - 2 \ep) (1 - \ep) \ep (1 - \la \zb)^2} \text{Bub}(s_{13})\times\\[5pt]
			&\quad \left[-\ep^2 \la n_f (1 - \la \zb)^2 + 
			  N_c (-3 (1 - (1 - \la) \zb) (1 - \la \zb)^2 + 
			  11 \ep (1 - (1 - \la) \zb) (1 - \la \zb)^2 \right.\\[5pt]
		        &\quad + \ep^3 \left(2 + \la - 2 \zb + 2 \la \zb + \la^3 \zb^2 (1 + 2 \zb) - 2 \la^2 \zb (3 - \zb + \zb^2)\right)  \\[5pt]
		        &\quad+ \left. \ep^2 (-9 (1 - \zb) - \la^3 \zb^2 (2 + 9 \zb) + 
		        \la^2 \zb (10 + 3 \zb + 9 \zb^2) - \la (2 - 3 \zb + 12 \zb^2)))\right]\\
		        &-\frac{N_c\zb} {(1 - \ep) \ep (1 - (1 - \la) \zb)} \text{Bub}(s_{23})	\times\\
		        & \left[(1 - (1 - \la) \zb)^2 - 3 \ep (1 - (1 - \la) \zb)^2 + 
  			\ep^2 (2 (1 - \zb)^2 - \la^2 (1 - 3 \zb) \zb + \la (1 + 4 \zb - 5 \zb^2)\right]\bigg\}
\end{align*}

\begin{align*}
A_{2c}^{(1)} &= c_\ep \bigg\{N_c 
			\left\{ \zb \,\text{Box}^{d+2}(s_{12},s_{13},s_{23})\times \right.\\ 
			&\quad\left[\la (1 - \zb - (-1 + \la) \la \zb^2) + \ep (1 - 3 \la (1 - \zb) - \zb - 2 \la^2 \zb^2 + 2 \la^3 \zb^2)\right]\\
		       &- \zb\,\text{Box}^{d+2}(s_{12},s_{23},s_{13})\times \\
		       &\quad\left[-(1 - \la) (1 - \zb - (-1 + \la) \la \zb^2) + \ep (2 - 2 \zb - 4 \la^2 \zb^2 + 2 \la^3 \zb^2 - \la (3 - 3 \zb - 2 \zb^2))\right] \\
		       &-\text{Box}^{d+2}(s_{13},s_{23},s_{12}) \times \\
		       &\left. \quad \left[1 - \zb - (-1 + \la) \la \zb^2 - \ep (2 - 2 \zb - 3 (-1 + \la) \la \zb^2 + (-1 + \la) \la \zb^3)\right]  \right\}\\
		       &+\frac{1}{(3 - 2 \ep) (1 - \ep) \ep}\text{Bub}(s_{12}) \times \\
		      &\quad \left[ \ep^2 (1 - \la) \la n_f \zb^2 +   N_c (3 (1 - \zb - (-1 + \la) \la \zb^2) - \right. \\
		      &\quad \,11 \ep (1 - \zb - (-1 + \la) \la \zb^2) + \ep^2 (9 - 9 \zb - \la (6 - 6 \zb - 11 \zb^2) +  \la^2 (6 - 6 \zb - 11 \zb^2)) + \\
		      & \quad \left.  \ep^3 (-2 (1 - \zb) + \la (4 - 4 \zb - 3 \zb^2) - \la^2 (4 - 4 \zb - 3 \zb^2)))\right]\\
		      &- \ep N_c \frac{(1 - \zb)}{(1 - \ep) (1 - (1 - \la) \zb) (1 - \la \zb)} \text{Bub}(m_h^2)\times\\
		      &\quad \left[(-(1 - \zb)^2 + 4 \la^3 \zb^2 - 2 \la^4 \zb^2 +   \la^2 (2 - 2 \zb - 5 \zb^2 + \zb^3) - \la (2 - 2 \zb - 3 \zb^2 + \zb^3) - \right.\\
		      &\quad \left.  2 \ep (1 - (1 + \la - \la^2) \zb - 2 (-1 + \la) \la \zb^2 - (-1 + \la)^2 \la^2 \zb^3)\right] \\
		      &+ N_c  \frac{1}{(1 - \ep) \ep (1 - \la \zb)}\text{Bub}(s_{13})\times\\
		      &\quad \left[(1 - (1 - \la) \zb) (1 - \la \zb)^2 -   3 \ep (1 - (1 - \la) \zb) (1 - \la \zb)^2 + \right.\\
		    &\quad \left.\ep^2 (2 - (1 + 3 \la) \zb - (1 - 6 \la + 3 \la^2) \zb^2 - \la (1 + \la - 2 \la^2) \zb^3)\right]\\
		    &- N_c \frac{1}{(1 - \ep) \ep (1 - (1 - \la) \zb)}\text{Bub}(s_{23})\times\\
		    &\quad \left[-(1 - (1 - \la) \zb)^2 (1 - \la \zb) + 3 \ep (1 - (1 - \la) \zb)^2 (1 - \la \zb) - \right.\\
		    &\quad\left.	  \ep^2 (2 - (4 - 3 \la) \zb - (-2 + 3 \la^2) \zb^2 - \la (3 - 5 \la + 2 \la^2) \zb^3)\right]\bigg\}
\end{align*}

\begin{align*}
A_{1}^{(1)} &= c_\ep \Bigg\{(1 - 2 \ep) N_c\bigg\{ 
		-\zb \, \text{Box}^{d+2}(s_{12},s_{13},s_{23}) \times\\
		&\quad  \left[\ep (1 - \la) (1 - \zb) - \la (1 - \zb - (-1 + \la - \la^2) \zb^2)\right]\\
		&+ \zb \left[1 - \zb + \zb^2 + 2 \la^2 \zb^2 - \la^3 \zb^2 - \la (1 - \ep (-1 + \zb) - \zb + 2 \zb^2)\right]\text{Box}^{d+2}(s_{12},s_{23},s_{13}) \\
		&- \left.\frac{1}{2}\left[2 - 2 \zb + 2 (1 - (1 + \ep) \la + (1 + \ep) \la^2) \zb^2 + 2 \ep (1 - \la) \la \zb^3\right]\text{Box}^{d+2}(s_{13},s_{23},s_{12}) \right\}\\
		& - \frac{1}{(3 - 2 \ep) (1 - \ep) \ep} \text{Bub}(s_{12})
		 \left[-\ep^2 (1 - \la) \la n_f \zb^2 \right.\\
		&- N_c \left(3 - 4 \ep^4 (-1 + \zb) - 3 \zb + 3 (1 - \la + \la^2) \zb^2 -  11 \ep (1 - \zb + (1 - \la + \la^2) \zb^2) -\right. \\
	      	&\quad\ep^3 (\la (4 - 4 \zb - 5 \zb^2) + 4 (3 - 3 \zb + \zb^2) + \la^2 (-4 + 4 \zb + 5 \zb^2)) - \\
      		&\,\, \left.\left.\ep^2 (\la^2 (6 - 6 \zb - 13 \zb^2) - 3 (5 - 5 \zb + 4 \zb^2) +  \la (-6 + 6 \zb + 13 \zb^2))\right)\right]\\
		&-\frac{\ep N_c (1 - \zb) }{(1 - \ep) (1 - (1 - \la) \zb)^2 (1 - \la \zb)^2}\text{Bub}(m_h^2)\times\\
		&\quad \left[(1 - \zb)^3 + 6 \la^5 \zb^4 - 2 \la^6 \zb^4 + \la^4 \zb^2 (4 - 4 \zb - 9 \zb^2 - \zb^3) - 2 \la^3 \zb^2 (4 - 4 \zb - 4 \zb^2 - \zb^3) +\right.\\
		&\quad  
		  \la (2 - 4 \zb + 8 \zb^2 - 8 \zb^3 + 4 \zb^4) - 
		  \la^2 (2 - 4 \zb + 4 \zb^2 - 4 \zb^3 + 7 \zb^4 + \zb^5) -  \\
		&\quad   2 \ep (2 - (5 - \la + \la^2) \zb +
	          5 \zb^2 - (2 + 4 \la - 6 \la^2 + 4 \la^3 - 2 \la^4) \zb^3 - \\
	         & \quad \left. 2 \la (-2 + 3 \la - 2 \la^2 + \la^3) \zb^4 - (-1 + \la)^2 \la^2 (2 - \la + \la^2) \zb^5)\right] \\
	          &-\frac{1}{(3 - 2 \ep) (1 - \ep) \ep (1 - \la \zb)^2}\text{Bub}(s_{13})\times\\
	          &\left\{(\ep^2 \la n_f \zb (1 - \la \zb)^2 + 
  		N_c \left(4 \ep^4 (1 - \la) (1 - \zb) \zb (1 - \la \zb)^2 -  \right.\right.\\
		 &\quad    3 (1 - \la \zb)^2 (1 - \zb + (1 - \la + \la^2) \zb^2) +  11 \ep (1 - \la \zb)^2 (1 - \zb + (1 - \la + \la^2) \zb^2) +\\
		 &\quad    \ep^3 (4 - (12 - \la) \zb + 2 (6 + 8 \la - 7 \la^2) \zb^2 - 
		  \la (28 - 16 \la - \la^2) \zb^3 + 4 \la^2 (3 - 3 \la + \la^2) \zb^4) -\\
		 & \quad\ep^2 (12 - 5 (3 + 4 \la) \zb + (15 + 21 \la + 10 \la^2) \zb^2 - 
		 \la (36 - 21 \la + 20 \la^2) \zb^3 + \\
	     &\quad \left. \left.   3 \la^2 (5 - 5 \la + 4 \la^2) \zb^4)\right)\right\}\\
	        &+\frac{1}{(3 - 2 \ep) (1 - \ep) \ep (1 - (1 - \la) \zb)^2}\text{Bub}(s_{23})\times\\
	        &\left\{-(1 - \la) n_f \zb (\ep - \ep (1 - \la) \zb)^2 + 
		  N_c \left[-4 \ep^4 \la (1 - \zb) \zb (1 - (1 - \la) \zb)^2 +  \right.\right.\\
		 &\quad 3 (1 - (1 - \la) \zb)^2 (1 - \zb + (1 - \la + \la^2) \zb^2) - \\
		 &\quad11 \ep (1 - (1 - \la) \zb)^2 (1 - \zb + (1 - \la + \la^2) \zb^2) - \\
		 &\quad \ep^3 \left(4 - (11 + \la) \zb - 2 (-7 - 6 \la + 7 \la^2) \zb^2 - (11 + 7 \la - 19 \la^2 + \la^3) \zb^3 + \right.\\
		 &\,\,\,\left.4 (-1 + \la)^2 (1 + \la + \la^2) \zb^4 \right) + 
		 \ep^2 (12 -  5 (7 - 4 \la) \zb + (46 - 41 \la + 10 \la^2) \zb^2 -  \\
		&\,\left.\left.(35 - 54 \la +  39 \la^2 - 20 \la^3) \zb^3 + 3 (1 - \la)^2 (4 - 3 \la + 4 \la^2) \zb^4)\right]\right\} \Bigg\}
\end{align*}

\subsection{Amplitude coefficients for $qg \to qh$}

\begin{align*}
A_{2}^{(1)}&=i\, c_\ep\left\{-\frac{N_c \left[\la - 2 \ep \la + \ep^2 (1 - \la) (1 - \zb)\right]}{1 - \la}\text{Box}^{d+2}(s_{12},s_{13},s_{23}) \right.\\
		& +\frac{\left[(1 - 2 \ep - \ep^2 (1 - \zb)\right]}{N_c}\text{Box}^{d+2}(s_{12},s_{23},s_{13}) \\
		&+\frac{N_c  \left[1 + \ep^2 (1 - \la) (1 - \zb) \zb -    \ep (2 - (-1 + \la) \zb + (-1 + \la) \zb^2)\right]}{(1 - \la) \zb}\text{Box}^{d+2}(s_{13},s_{23},s_{12}) \\
		& +\frac{1}{2\ep N_c}\left(\ep^2 - \frac{N_c^2 (2 - 4 \ep - \ep^2 (1 - \la) \zb)}{(1 - \la) \zb}\right) \text{Bub}(s_{12})\\
		&+\frac{\ep (1 - \zb)}{(1 - \ep) (1 - \la)  \zb (1 - (1 - \la) \zb) (1 - \la \zb)^2 \,N_c} \text{Bub}(m_h^2) \times \\
		&\quad \left[-\ep (1 - \la \zb) (-(1 - \la) \zb (1 - \la \zb) + \right.\\
   		&\quad\left.  N_c^2 (2 - (1 - \la) \zb - (1 - \la) \la \zb^2)) + (1 - \la) \zb (-(1 - \la \zb)^2 + N_c^2 (3 - 2 \zb + \la^2 \zb^2))\right] \\
		&+\frac{1}{2(3 - 2 \ep) (1 - \ep) \ep (1 - \la) N_c \zb (1 - \la \zb)^2}\text{Bub}(s_{13})\times\\
		&\quad \left[6 (1 - \la \zb)^2 + 4 \ep^4 (1 - \la \zb)^2 -   \ep (13 - 13 N_c^2 + 4 N_c n_f) (1 - \la \zb)^2 + \right.\\
		&\quad  \ep^2 (15 (1 - \la \zb)^2 + 8 N_c n_f (1 - \la \zb)^2 -  3 N_c^2 (7 + (4 - 18 \la) \zb - (4 - 4 \la - 7 \la^2) \zb^2)) + \\
		&\quad \left. 4 \ep^3 (-3 (1 - \la \zb)^2 - N_c n_f (1 - \la \zb)^2 +  2 N_c^2 (1 + \zb - 3 \la \zb - (1 - \la - \la^2) \zb^2))\right] \\
		&-\frac{1}{\ep (1 - \la) N_c \zb (1 - (1 - \la) \zb)}\text{Bub}(s_{23})\times\\
		&\left.\quad \left[-\ep^2 \zb (1 - \la - \zb + \la \zb) +   N_c^2 (1 - (1 - \la) \zb + \ep^2 (1 - \la) (1 - \zb) \zb -      2 \ep (1 - (1 - \la) \zb))\right]\right.\Bigg\}
\end{align*}

\begin{align*}
A_{3}^{(1)} &=i\, c_\ep\left\{- \frac{N_c\left[\ep^2 (1 - \la) (1 - \zb) + \la^2 \zb -  \ep (1 - \la - \zb + \la \zb + 2 \la^2 \zb)\right]}{(1 - \la) \la \zb}\text{Box}^{d+2}(s_{12},s_{13},s_{23})\right. \\
		&-\frac{\left[-\ep^2 (1 - \zb) - \la \zb + 2 \ep \la \zb\right]}{\la N_c \zb}\text{Box}^{d+2}(s_{12},s_{23},s_{13}) \\
		&+\frac{ N_c (\la - 2 \ep \la + \ep^2 (1 - \la) (1 - \zb))}{(1 - \la) \la \zb}\text{Box}^{d+2}(s_{13},s_{23},s_{12}) \\
		&-\frac{1}{\ep (1 - \la) \la N_c \zb^2} \left[-\ep^2 (1 - \la) (1 - N_c^2) (1 - \zb) + \la N_c^2 \zb - 2 \ep \la N_c^2 \zb\right]\text{Bub}(s_{12})\\
		& +\frac{\ep (1 - \zb)}{(1 - \ep) (1 - \la) \la N_c \zb^2 (1 - \la \zb)^2} \text{Bub}(m_h^2)
		 \left[\ep (1 - \la \zb) (1 - N_c^2 - \la^2 (-1 + N_c^2) \zb - \right.\\
		&\quad  \left.   \la (1 + N_c^2 (-1 + \zb) + \zb)) + (1 - \la) (-(1 - \la \zb)^2 +  N_c^2 (1 - 2 \la (1 - \zb) \zb + \la^2 \zb^2))\right]\\
		&+\frac{1}{2(3 - 2 \ep) (1 - \ep) \ep (1 - \la) N_c \zb (1 - \la \zb)^2}\text{Bub}(s_{13}) \times\\
		&\quad\left[6 (1 - \la \zb)^2 + 4 \ep^4 (1 - \la \zb)^2 -  \ep (13 - 13 N_c^2 + 4 N_c n_f) (1 - \la \zb)^2 + \right.\\
		&\quad\, \ep^2 \left(15 (1 - \la \zb)^2 + 8 N_c n_f (1 - \la \zb)^2 -  3 N_c^2 (7 + (4 - 18 \la) \zb - (4 - 4 \la - 7 \la^2) \zb^2)\right) + \\
		&\quad\left. 4 \ep^3 \left(-3 (1 - \la \zb)^2 - N_c n_f (1 - \la \zb)^2 + 2 N_c^2 (1 + \zb - 3 \la \zb - (1 - \la - \la^2) \zb^2)\right)\right]\\[5 pt]
		&+\left.\frac{(-2 \la N_c^2 + 4 \ep \la N_c^2 - \ep^2 (1 - \la) (1 + N_c^2))}{2\ep (1 - \la) \la N_c \zb}\text{Bub}(s_{23})\right\}
\end{align*}

\subsection{Amplitude coefficients for $q\bar q \to gh$}

\begin{align*}
A_{2}^{(1)}&=i\, c_\ep\left\{\frac{ -N_c \left[-\ep^2 (1 - \zb) - (1 - \la) \la \zb^2 +    2 \ep (1 - \la) \la \zb^2\right]}{(1 - \la) \zb}\text{Box}^{d+2}(s_{12},s_{13},s_{23}) \right.\\
		&+\frac{ N_c \left[-\ep^2 (1 - \zb) + (1 - \la)^2 \zb^2 +    \ep (1 - \zb - 2 (-1 + \la)^2 \zb^2)\right]}{(1 - \la) \zb}\text{Box}^{d+2}(s_{12},s_{23},s_{13}) \\
		&-\frac{ \left[\ep^2 (1 - \zb) + \zb - 2 \ep (1 - \la) \zb - \la \zb\right]}{(1 - \la) N_c \zb}\text{Box}^{d+2}(s_{13},s_{23},s_{12}) \\
		&-\frac{1 }{2(3 - 2 \ep) (1 - \ep) \ep N_c \zb^2}\,\text{Bub}(s_{12})
		\left[6 \zb^2 + 4 \ep^4 \zb^2 - \ep (13 - 13 N_c^2 + 4 N_c n_f) \zb^2 + \right.\\
	       & \left. \ep^2 (15 \zb^2 + 8 N_c n_f \zb^2 + 3 N_c^2 (4 - 4 \zb - 7 \zb^2)) + 
	       4 \ep^3 (-3 \zb^2 - N_c n_f \zb^2 - 2 N_c^2 (1 - \zb - \zb^2))\right]\\[5 pt]
	       &+\frac{ \ep  (1 - \zb)}{(1 - \ep) (1 - \la) N_c \zb^2 (1 - (1 - \la) \zb)}\text{Bub}(m_h^2) \times\\[5 pt]
	       &\quad\left[\zb - \ep \zb +   N_c^2 (2 - (3 + \ep) \zb + 2 \ep \zb^2 - 2 \la (1 - 2 \zb) (1 - \ep \zb) -      2 \la^2 \zb (1 - \ep \zb))\right] \\
	       &-\frac{1}{2\ep (1 - \la) N_c \zb}\left[ \ep^2 (1 + N_c^2) - 2 (1 - \la) N_c^2 \zb + 4 \ep (1 - \la) N_c^2 \zb\right]\text{Bub}(s_{13})\\
	       &\left.+\frac{1}{N_c}\left[-2 N_c^2 + \frac{N_c^2}{\ep} - \frac{\ep (1 - N_c^2) (1 - \zb)}{(1 - \la) \zb (1 - (1 - \la) \zb)}\right]\text{Bub}(s_{23})\right\}
\end{align*}

\begin{align*}
A_{3}^{(1)} &= i\, c_\ep\left\{\frac{N_c \left[-\ep^2 (1 - \zb) + \la^2 \zb^2 + \ep (1 - \zb - 2 \la^2 \zb^2)\right]}{\la \zb} \text{Box}^{d+2}(s_{12},s_{13},s_{23}) \right.\\
		 &+\frac{ N_c \left[\ep^2 (1 - \zb) + (1 - \la) \la \zb^2 - 2 \ep (1 - \la) \la \zb^2\right]}{\la \zb}\text{Box}^{d+2}(s_{12},s_{23},s_{13}) \\
		 &-\frac{ \left[-\ep^2 (1 - \zb) - \la \zb + 2 \ep \la \zb\right]}{\la N_c \zb}\text{Box}^{d+2}(s_{13},s_{23},s_{12}) \\
		 &-\frac{1}{2(3 - 2 \ep) (1 - \ep) \ep N_c \zb^2}\, \text{Bub}(s_{12})
		 \left[6 \zb^2 + 4 \ep^4 \zb^2 - \ep (13 - 13 N_c^2 + 4 N_c n_f) \zb^2 + \right.\\
  		 &\left.\ep^2 (15 \zb^2 + 8 N_c n_f \zb^2 + 3 N_c^2 (4 - 4 \zb - 7 \zb^2)) + 
		 4 \ep^3 (-3 \zb^2 - N_c n_f \zb^2 - 2 N_c^2 (1 - \zb - \zb^2))\right] \\[5 pt]
		 &+\frac{\ep  (1 - \zb)}{(1 - \ep) \la N_c \zb^2 (1 - \la \zb)} \text{Bub}(m_h^2) \times\\
		 &\quad\left[(1 - \ep) (1 - N_c^2) \zb +    2 \la N_c^2 (1 - \ep \zb) - 2 \la^2 N_c^2 \zb (1 - \ep \zb)\right]\\[5 pt]
		 &+\frac{1 }{N_c}\left[-2 N_c^2+\frac{N_c^2}{\ep}-\frac{(\ep (1-N_c^2) (1-\zb))}{(\la \zb (1-\la \zb))}\right]\text{Bub}(s_{13})\\[5 pt]
		 &-\left.\frac{ (\ep^2 (1 + N_c^2) - 2 \la N_c^2 \zb + 4 \ep \la N_c^2 \zb) }{2\ep \la N_c \zb}\text{Bub}(s_{23})\right\}
\end{align*}



\bibliography{biblio}


\end{document}